# Few-shot Class-variable Incremental Audio Classification via Prototype Adaptation and Pseudo Class-variable Training

*Yanxiong Li, Guoqing Chen, Qianqian Li, Sen Huang**

School of Electronic and Information Engineering, South China University of Technology, Guangzhou, China
eeyxli@scut.edu.cn, cgq2971@gmail.com, 1210517956@qq.com, huangsen@scut.edu.cn

## Abstract

In the task of few-shot class-incremental audio classification, the number of classes is assumed to always increase without considering the possibility of decrease. However, the number of classes generally increases or decreases in practice. In this paper, we investigate a problem of Few-shot Class-variable Incremental Audio Classification (FCIAC), where the number of classes increases or decreases. We propose a FCIAC method using prototype adaptation and pseudo class-variable training. The model in our method consists of an encoder and a classifier. The classifier is initialized by a class-variable prototype adaptation network, whose structure dynamically changes with the change of classes. In addition, we design a pseudo class-variable training strategy to enhance the model's adaptability to changing classes. Experiments on three public datasets show that our method exceeds previous methods in average accuracy. The code is at: https://github.com/cgq2971-afk/FCIAC.

## 1. Introduction

Audio classification is a key technique for many multimedia processing tasks, including audio content analysis [1], domestic activity estimation [2], acoustic scene classification [3]-[5], bio-acoustic monitoring [6], [7], audio-visual video understanding [8], [9], road audio surveillance [10], and speaker diarization [11]. Conventional supervised audio classification methods rely on abundant labeled data [12]-[16] and can recognize the predefined classes only. To recognize old classes (classes appeared in prior sessions) and new classes (classes appeared in current sessions), the model needs to be retrained with the samples of both old and new classes. Privacy protection and copyright restrictions often prevent the samples of old classes (samples previously used) from being accessed for common users. To recognize old and new classes, the model needs to be retrained with their samples. Few-shot audio classification methods [17]-[20] relieve the model's demand for training samples, in which the models tend to overfit new classes.

In recent years, the Few-shot Class-incremental Audio Classification (FCAC) methods are proposed to memorize old classes and continually recognize new classes using a few samples [21]-[28]. The models in the FCAC methods are generally composed of an encoder and a classifier. The encoders are basically a convolutional neural network, whereas the classifiers are different in structure. The main classifiers used in the FCAC methods include the ridge regression classifier [21], prototype classifier [22]-[24], stochastic classifier [25], [26], and linear classifier [27], [28]. Although current FCAC methods can continually recognize new classes without forgetting old ones, they assume that the number of classes always increases. The above assumption contradicts the requirements of many practical applications where the number of classes will both increase and decrease. For example, the users of smart speakers hope that the voice keywords can be added (increasing the number of classes) and removed (decreasing the number of classes), rather than constantly adding voice keywords. To overcome the above limitation of the FCAC methods, we define a new problem of FCIAC and propose an effective FCIAC method. In the proposed method, the number of classes either increases or decreases in each incremental session. Experiments on three public datasets (LS-100, NSynth-100, and FSC-89) show that our method exceeds the baseline methods in average accuracy. Main contributions of this work are as follows.

1) We define a FCIAC problem where the number of classes either increases or decreases in incremental sessions.

2) We propose a FCIAC method whose model consists of an encoder and a classifier. The encoder is trained in the base session and frozen in the incremental sessions. The classifier is updated in each incremental session by a Class-variable Prototype Adaptation Network (CPAN) for enhancing the model's ability to distinguish between various classes.

3) We design a Pseudo Class-variable Training Strategy (PCTS) in the base session for simulating the model training in the incremental sessions, thereby enhancing the model's adaptability to the change of classes.

## 2. Method

In this section, we describe the details of our method, including the problem definition, method framework, model architecture, PCTS, and embedding reconstruction.

### 2.1. Problem Definition

The FCIAC aims to continually recognize new classes using few training samples without forgetting the old ones, where the classes can be added or removed in each incremental session.

The training and evaluation data of different sessions are denoted as $\{\boldsymbol{D}_0^t, \boldsymbol{D}_1^t, \cdots, \boldsymbol{D}_m^t, \boldsymbol{D}_M^t\}$ and $\{\boldsymbol{D}_0^e, \boldsymbol{D}_1^e, \cdots, \boldsymbol{D}_m^e, \boldsymbol{D}_M^e\}$, respectively, where *t*, *e*, and *M* denote training, evaluation, and the total number of sessions, respectively. For each session *m*, $\boldsymbol{D}_m^t$ and $\boldsymbol{D}_m^e$ share the same label set $\boldsymbol{L}_m$. The datasets in different sessions do not have the same kind of classes, i.e., $\forall i, j$ and $i \neq j, L_i \cap L_j = \emptyset$ where ø denotes an empty set. If the classes are removed in the *m*th ($m \geq 1$) session, there are no new training and testing samples in the *m*th session, namely $\boldsymbol{D}_m^t = \emptyset$ and $\boldsymbol{D}_m^e = \emptyset$. In the *m*th session, only $\boldsymbol{D}_m^t$ can be used to train the model, and the trained model is evaluated on the evaluation datasets of both current and all prior sessions, excluding the classes that have been removed. Session 0 is the base session with a large-scale dataset $\boldsymbol{D}_0^t$ to train the model. The dataset $\boldsymbol{D}_m^t$ ($m \geqslant 1$) is a $N$-way $K$-shot training (small-scale) dataset if classes are added in the $m$th session; otherwise, it is an empty set. The classes appeared in base and incremental sessions are called base and incremental classes, respectively.

* Corresponding author: Sen Huang.

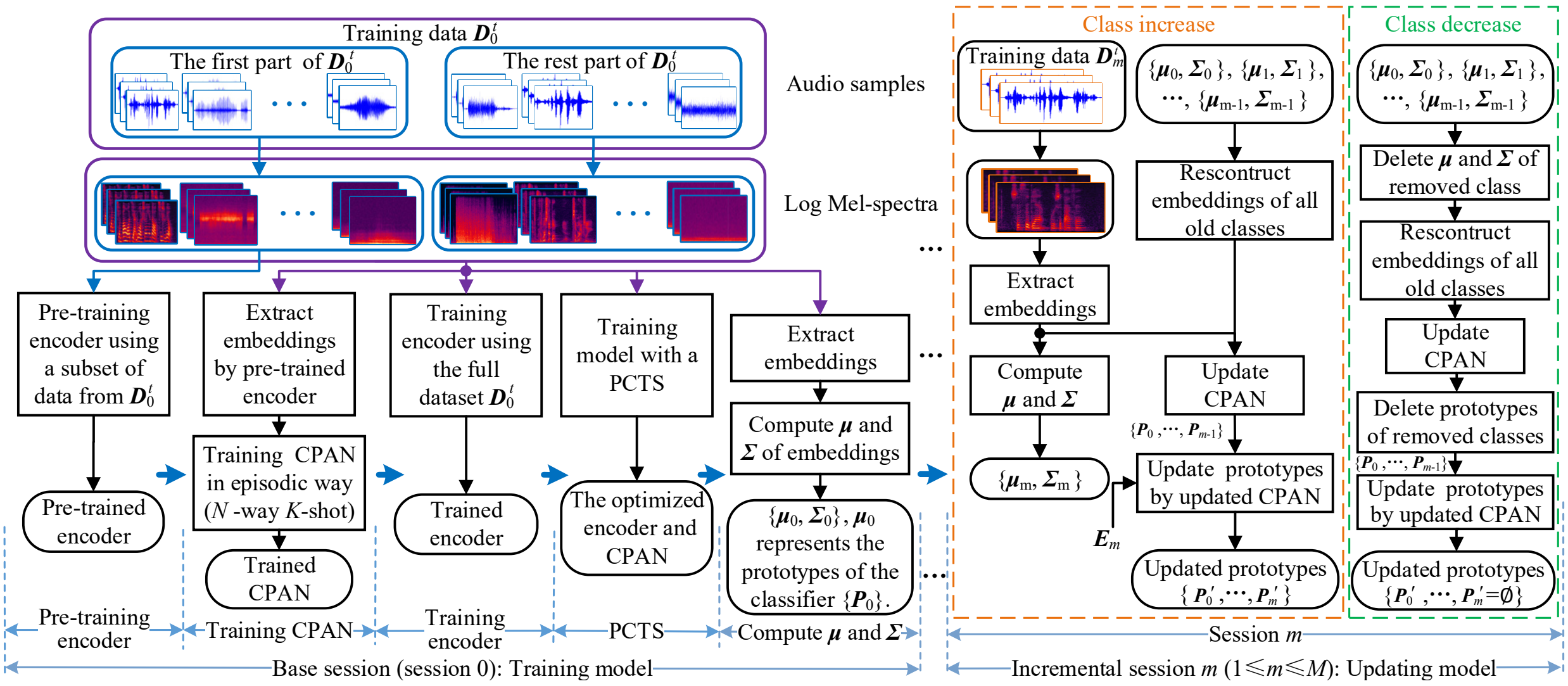

Figure 1: *Framework for the proposed FCIAC method. The encoder, CPAN and classifier are sequentially trained in base session, while the CPAN and prototypes of classifier are updated in incremental sessions. In the incremental sessions, when the number of classes increases, the operations in the orange dashed box are executed; Otherwise, the operation in the green dashed box are executed. CPAN: Class-variable Prototype Adaptation Network; PCTS: Pseudo Class-variable (PCV) Training Strategy;* $\boldsymbol{E}$*: embeddings extracted by encoder;* $\boldsymbol{\mu}$*: mean vector;* $\boldsymbol{\Sigma}$*: covariance matrix;* $\boldsymbol{P}$*: prototypes.*

## 2.2. Method Framework

As illustrated in Fig. 1, the proposed FCIAC framework consists of a base session and $M$ incremental sessions. The base session has five steps, including pre-training the encoder, training the CPAN, training the encoder, training the model with PCTS, and computing the mean vectors and covariance matrices of embeddings for base classes. The mean vectors are used as prototypes to initialize the classifier. Both the mean vectors and covariance matrices are saved to reconstruct the embeddings of old classes in next incremental sessions. To mitigate the model's forgetting of base classes, the encoder is frozen after training in the base session. To enhance the model's plasticity to incremental classes, the CPAN is updated in each incremental session. When the number of classes increases in current session, the embeddings of old classes are reconstructed using their corresponding mean vectors and covariance matrices. The embeddings of new and old classes are used to update the prototypes. Meanwhile, the mean vectors and covariance matrices of the new class's embeddings are computed and stored. When some classes are removed in current session, the prototypes, mean vectors and covariance matrices of the removed classes are discarded, and the prototypes are updated using the reconstructed embeddings of the existing (unremoved) classes. This design allows the model to flexibly adapt to both class addition and removal, and to keep its discriminability to the existing classes.

## 2.3. Model Architecture

The ResNet [29] is proved to be effective for learning discriminative embeddings in many applications [30]-[32]. Motivated by its success, a ResNet-based network is used as encoder. The encoder adopts the default parameters of ResNet-18. The classifier consists of prototypes and each prototype is used to represent one class. To ensure that the classifier maintains discriminative decision boundaries for all classes after adding or removing classes, we design a class-variable prototype adaptation network that dynamically adjusts prototypes in response to both class addition and removal during incremental sessions. As shown in Fig. 2, the proposed CPAN is composed of four modules, including Attentive Prototype Generator Module (APGM), Stability Adaptation Module for Prototypes (SAMP), Plasticity Adaptation Module for Prototypes (PAMP), and a fusion module.

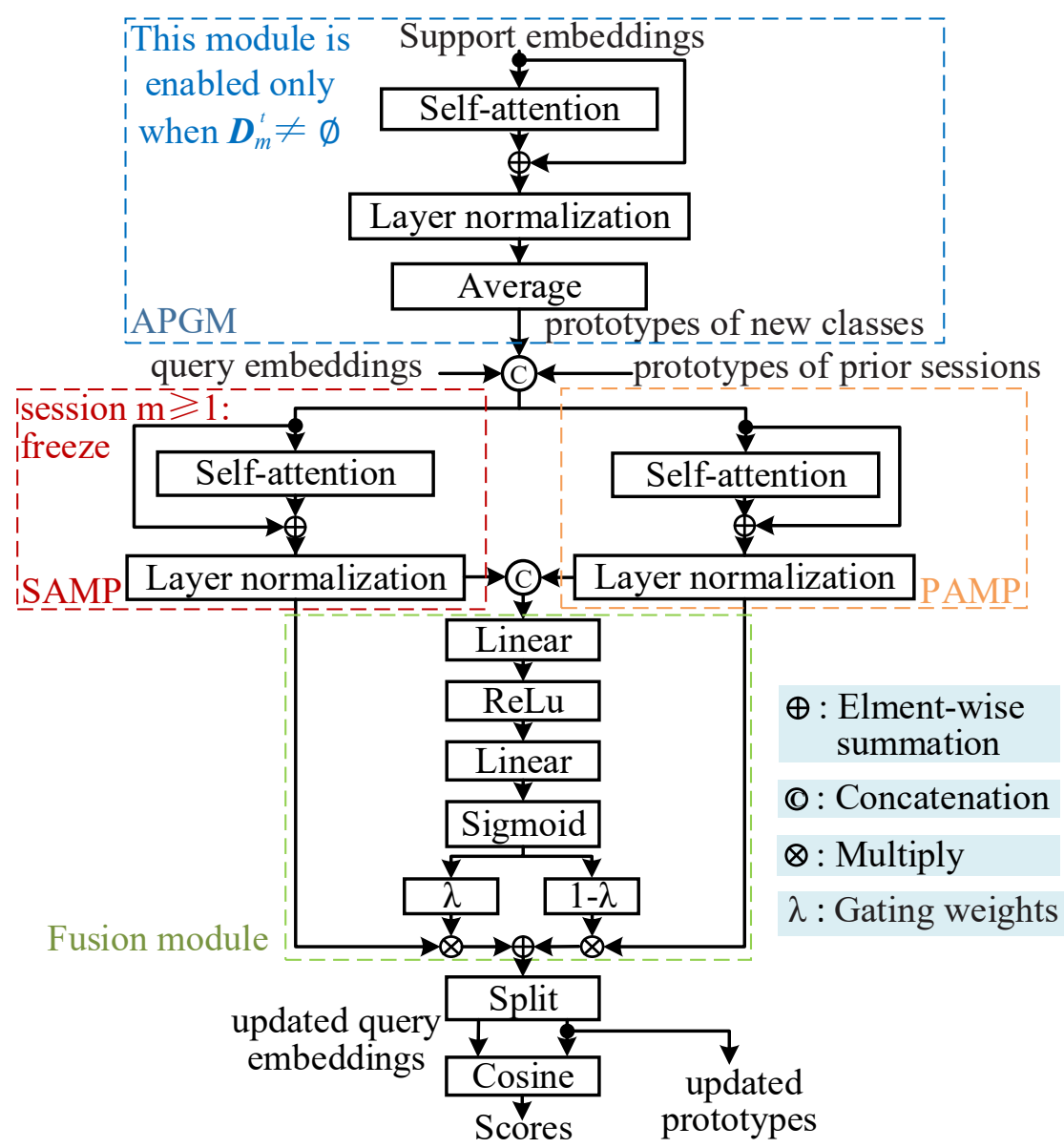

Figure 2: *Architecture of class-variable prototype adaptation network.*

The APGM is activated only during class addition in the incremental session and is used to generate prototypes from the support embeddings of the current session. Both the SAMP and PAMP modules are used to update prototypes and query embeddings. Their inputs include the generated prototypes (which are absent in the session of class removal), the current session's query embeddings, and the concatenated results of the previous session's prototypes. The fusion module combines the outputs from SAMP and PAMP to obtain the final prototype updates. The design of CPAN is motivated by the need to effectively utilize limited data to generate representative prototypes for new classes and to adjust the prototypes of all classes in the current session. The self-attention mechanism [33] is adopted in the APGM, SAMP, and PAMP modules, as it captures the relationships among all input vectors. Each module mainly consists of self-attention, layer normalization [34], and residual connections. In APGM, self-attention captures the relationships among support embeddings of new classes, enabling representative prototypes that enhance the discriminative power of the prototype classifier. In SAMP and PAMP, self-attention captures global positional information of all prototypes in the prototype space and adjusts them accordingly. While SAMP is frozen after training in the base session, PAMP continues to update during the incremental session using support samples from new classes and reconstructed embeddings from old classes. The fusion module then combines their outputs to update the prototypes. Finally, cosine similarity between each updated query embedding and the updated prototypes is used for classification.

### 2.4. Pseudo Class-variable Training Strategy

In FCIAC, the number of samples per incremental class is limited, and the number of classes either increases or decreases across incremental sessions. However, during the base session, there is sufficient training data, and the model is restricted to recognizing predefined base classes. To make the model trained in the base session to dynamically-changing incremental classes without using the samples of incremental classes, we propose a PCTS. First, we construct multiple episodes, and each episode contains $N{\cdot}K$ samples ($N$ classes and $K$ samples per class) synthesized by the training data of base classes. Then, we alternately train model for class addition and class removal on the samples of each episode. The above steps enable the model to recognize new classes and memorize existing classes. The PCTS is summarized as Algorithm 1, as shown below.

**Algorithm 1: Pseudo Class-Variable Training Strategy**

**Input**: Base-class training set $\boldsymbol{D}_0^t$; Encoder; CPAN.

**Do:**

1. Randomly select two subsets from the $\boldsymbol{D}_0^t$ , which will be used to synthesize pseudo-classes.
2. Sample a mixing coefficient $\gamma \sim Beta(a, a)$, where $Beta(a, a)$ denotes a symmetric Beta distribution on [0,1] (centered at 0.5), and generate pseudo-classes samples by mixing up two subsets. Each sample of pseudo-classes is assigned a label.
3. Construct pseudo-class dataset $\boldsymbol{D}_v$ and split $\boldsymbol{D}_v$ into a support set $\boldsymbol{S}_v$ and a query set $\boldsymbol{Q}_v$. Use encoder to learn embeddings from $\boldsymbol{S}_v$ and $\boldsymbol{Q}_v$.
4. Use the APGM to produce the prototypes of current session.
5. Freeze the SAMP, and activate both the PAMP and fusion module.
6. Update all prototypes and adjust the query embeddings via the CPAN.
7. Predict the query labels and compute the cross-entropy loss $\mathcal{L}_{\text{CE}}$.
8. Update encoder and CPAN by stochastic gradient descent algorithm.
9. Freeze encoder to memorize the learned knowledge of old classes.
10. Randomly discard few prototypes of old classes.
11. Update the remaining classes' prototypes and their query embeddings via the CPAN.
12. Predict the query labels of the existing-classes' and compute the $\mathcal{L}_{\text{CE}}$.
13. Update the CPAN by stochastic gradient descent algorithm.
14. Repeat the above steps until reaching the maximum training cycles.

**Output**: The optimized encoder and CPAN.

### 2.5. Embeddings Reconstruction

The embeddings of the same class $E = \{\boldsymbol{e_k}\}(1 \leq k \leq K)$ are assumed to follow a multivariate Gaussian distribution $\mathcal{N}(\boldsymbol{\mu}, \boldsymbol{\Sigma})$. The probability density function is defined by:

$$P(\boldsymbol{e_k}) = \frac{1}{(2\pi)^{D/2} \mid \boldsymbol{\Sigma} \mid^{1/2}} \exp\left(-\frac{1}{2}(\boldsymbol{e_k} - \boldsymbol{\mu})^T \boldsymbol{\Sigma}^{-1} (\boldsymbol{e_k} - \boldsymbol{\mu})\right) \quad (1)$$

where $D$ is the dimension of the embedding $\boldsymbol{e_k}$ . The multivariate Gaussian distribution is used to represent each class and can also be used to reconstruct other embeddings of the same class. The process of embedding reconstruction is as follows. First, a random vector $\boldsymbol{\epsilon}$ is generated from a standard normal distribution. The dimension of this vector is the same as the mean vector $\boldsymbol{\mu}$ of each class. Then, the random vector $\boldsymbol{\epsilon}$ is multiplied by the inverse of the covariance matrix $\boldsymbol{\Sigma}^{-1}$, and the result is added to the mean vector $\boldsymbol{\mu}$ to obtain the reconstructed embedding. This reconstruction process is represented by:

$$\boldsymbol{e}_{\boldsymbol{k}}' = \boldsymbol{\mu} + \boldsymbol{\epsilon}\boldsymbol{\Sigma}^{-1} \quad (2)$$

## 3. Experiments

This section gives our experiments, including experimental datasets and setup, comparison and ablation experiments.

### 3.1. Experimental Datasets

Experiments are done on three public datasets, including the LS-100, NSynth-100, and FSC-89. LS-100 consists of speech utterances from 100 different speakers. NSynth-100 comprises 100 classes of musical instrument sounds. FSC-89 contains 89 sound event classes spanning a broad range of real-world sounds, including human and animal sounds, as well as natural, musical, and miscellaneous categories.

They are widely used in prior studies, whose details are presented in [21]. The three datasets can be downloaded from https://www.modelscope.cn/profile/pp199124903.

### 3.2. Experimental Setup

Average Accuracy (AA) is used to assess the overall performance of different methods and is defined by

$$AA = \frac{1}{M} \sum_{m=0}^{M-1} A^m \quad (3)$$

where $A^m$ stands for the accuracy in session $m$. Across all three datasets, we configure 4 incremental sessions with an alternating schedule: each session adds 5 new classes and removes 2 old classes (one base class and one previously added incremental class). The value of $N$-way $K$-shot for few-shot learning is set to 5-way 5-shot. The testing process is repeated 100 times, and we report the AA. The dimensions of the log-Mel spectrogram and prototype are set to 128 and 512, respectively. In the base session, the encoder and CPAN are trained with a learning rate of 0.1 and $2 \times 10^{-4}$, respectively. In incremental sessions, all learning rates are set to $1 \times 10^{-4}$.

### 3.3. Comparison of Different Methods

In this subsection, we compare our method to state-of-the-art methods, including the CEC [35], PAN [22] and AMFO [28]. All methods are configured according to the FCIAC setting and compared under the same conditions. Tables 1, 2 and 3 list the results obtained by various methods for different classes (base, incremental and all classes) on the LS-100, NSynth-100 and FSC-89, respectively. In the three tables, "1 (+5)" denotes that 5 classes are added in session 1, and "2 (-2)" denotes that 2 classes are removed in session 2. The results show that our method almost obtains the highest AA scores among all methods for the base, incremental and all classes on all datasets.

Table 1: *Results obtained by various methods on the LS-100.*

| Method | Session | Accuracy in various sessions (%) 0 | 1 (+5) | 2 (-2) | 3 (+5) | 4 (-2) | AA (%) |
|---|---|---|---|---|---|---|---|
| CEC [35] | Base | 91.69 | 91.94 | 91.53 | 91.48 | 91.35 | 91.60 |
| | Inc. | - | 85.53 | 86.38 | 80.65 | 83.43 | 84.00 |
| | All | 91.69 | 91.45 | 91.20 | 90.05 | 90.39 | 90.96 |
| PAN [22] | Base | 91.80 | 91.74 | 91.66 | 91.61 | 91.46 | 91.65 |
| | Inc. | - | 88.37 | 90.25 | 89.53 | 87.25 | 88.85 |
| | All | 91.80 | 91.48 | 91.57 | 91.33 | 90.95 | 91.43 |
| AMFO [28] | Base | 92.28 | 92.23 | 91.98 | **91.73** | 91.31 | 91.91 |
| | Inc. | - | 93.00 | 95.00 | 82.22 | 84.38 | 88.65 |
| | All | 92.28 | 92.29 | 92.17 | 90.45 | 90.47 | 91.53 |
| Ours | Base | **92.60** | **92.45** | **92.39** | 91.65 | **91.70** | **92.16** |
| | Inc. | - | **98.82** | **97.75** | **97.32** | **97.75** | **97.91** |
| | All | **92.60** | **92.94** | **92.73** | **92.40** | **92.41** | **92.62** |

Table 2: *Results obtained by various methods on the NSynth-100.*

| Method | Session | Accuracy in various sessions (%) 0 | 1 (+5) | 2 (-2) | 3 (+5) | 4 (-2) | AA (%) |
|---|---|---|---|---|---|---|---|
| CEC [35] | Base | 99.91 | 99.79 | 99.80 | 99.73 | 99.89 | 99.82 |
| | Inc. | - | 66.53 | 65.91 | 69.95 | 67.59 | 67.50 |
| | All | 99.91 | 97.02 | 97.46 | 95.48 | 95.65 | 97.10 |
| PAN [22] | Base | **99.93** | **99.87** | 99.82 | 99.80 | 99.83 | **99.85** |
| | Inc. | - | 75.86 | 74.75 | 77.82 | 78.00 | 76.61 |
| | All | **99.93** | 97.87 | 98.09 | 96.66 | 96.97 | 97.90 |
| AMFO [28] | Base | 99.85 | 99.82 | 99.81 | 99.47 | 99.01 | 99.59 |
| | Inc. | - | 80.53 | 79.26 | 75.17 | 76.36 | 77.83 |
| | All | 99.85 | 98.21 | 98.39 | 96.00 | 96.04 | 97.70 |
| Ours | Base | 99.91 | 99.84 | **99.87** | **99.84** | 99.81 | **99.85** |
| | Inc. | - | **86.42** | **84.50** | **86.92** | **87.50** | **86.34** |
| | All | 99.91 | **98.72** | **98.81** | **98.00** | **98.20** | **98.73** |

Table 3: *Results obtained by various methods on the FSC-89.*

| Method | Session | Accuracy in various sessions (%) 0 | 1 (+5) | 2 (-2) | 3 (+5) | 4 (-2) | AA (%) |
|---|---|---|---|---|---|---|---|
| CEC [35] | Base | 42.57 | 40.89 | 40.03 | 39.16 | 39.34 | 40.40 |
| | Inc. | - | 22.96 | 21.35 | 19.93 | 20.13 | 21.09 |
| | All | 42.57 | 39.49 | 38.82 | 36.58 | 36.98 | 38.89 |
| PAN [22] | Base | 43.68 | 41.90 | 41.00 | 40.41 | 40.58 | 41.51 |
| | Inc. | - | 29.97 | 30.38 | 22.70 | **28.19** | 27.81 |
| | All | 43.68 | 40.96 | 40.31 | 38.03 | **39.06** | 40.41 |
| AMFO [28] | Base | **43.97** | **41.99** | 41.13 | 40.53 | 40.11 | 41.55 |
| | Inc. | - | 32.10 | 30.15 | 24.13 | 25.36 | 27.94 |
| | All | **43.97** | 41.21 | 40.42 | 38.33 | 38.30 | 40.45 |
| Ours | Base | 43.88 | 41.92 | **41.27** | **40.79** | **40.85** | **41.74** |
| | Inc. | - | **35.83** | **33.88** | **24.71** | 26.19 | **30.15** |
| | All | 43.88 | **41.44** | **40.79** | **38.63** | 39.05 | **40.76** |

Then, we consider more realistic deployment scenarios for intelligent audio terminals, where the class set may evolve in an irregular and unpredictable manner. we design two more challenging settings on LS-100. In the first setting, the numbers of added classes and removed classes are random in each incremental session. In the second setting, class addition and class removal occur simultaneously within the same session, which better reflects real-world applications (new keywords may be introduced while obsolete ones are removed at the same time). For ease of analysis, all experiments are done on LS-100 with four incremental sessions, and the AA scores of all classes in each session are reported in Figure 3. In Figure 3 (b), "(-1+7)" denotes that 1 class is removed and 7 classes are added. As shown in Figure 3, our method consistently achieves the highest AA scores in each session. In addition, the deviations of its AA scores are small, which shows that the variation of class number in each session has a small impact on the accuracy of our method.

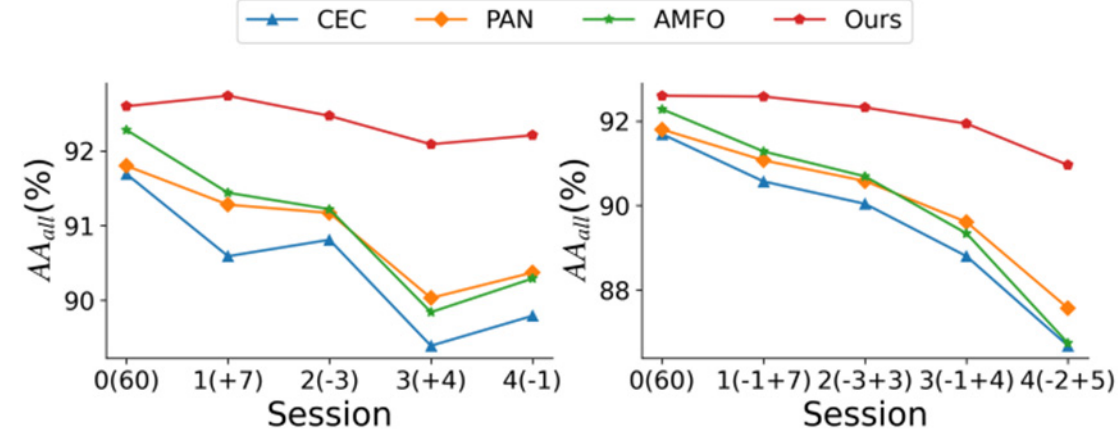

Figure 3: *The AA scores of all classes in each session achieved by different methods under the two class-dynamic settings on LS-100.*

We further evaluate whether the differences among various methods are statistically significant. Following common practice [24]-[28], we apply the Friedman test [36] and conduct post-hoc analysis using the Nemenyi test [37]. The tests are conducted on the average ranks computed from 10 runs on each experimental dataset. Figure 4 (a) reports how often each method attains each rank, while Figure 4 (b) shows the significance results at confidence level $\alpha = 0.05$. In Figure 4 (b), digits 1 to 4 denote the mean ranks, where smaller values indicate better performance. Methods connected by thick black crossbars show no statistically significant difference. As shown in Figure 4, our approach outperforms all baselines in accuracy, and the improvement is statistically significant.

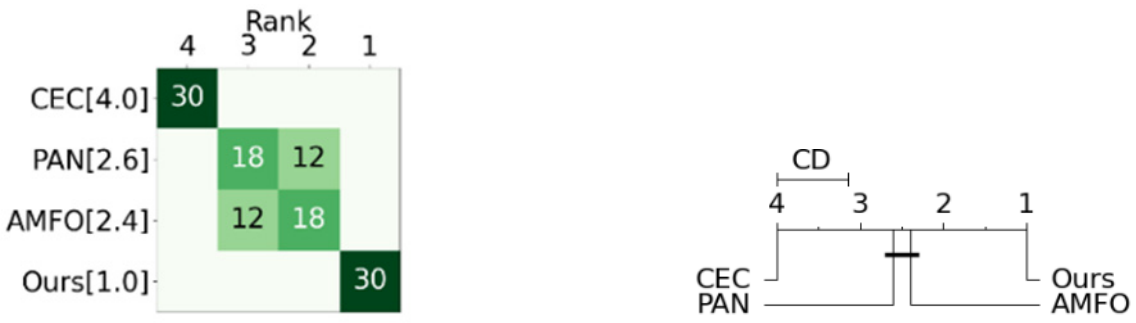

(a) *Histogram of method ranks* (b) *Critical difference (CD) diagram*

Figure 4: *Results of significance test*. (a) *Visualization of the number of times each method achieves each rank*. (b) *Visualization of significance differences between different methods.*

### 3.4. Ablation Experiments

In this subsection, we conduct an ablation study on the LS-100 to assess the contributions of the CPAN and PCTS to the performance of our method. As shown in Table 4, the CPAN and PCTS have contributions to the improvement of AA scores obtained by our method. When they are used, our method obtains the highest AA score of 92.62% for all classes.

Table 4: *The AA scores (in %) obtained by our method with various combinations of CAPAN and PCTS on the LS-100.*

| CPAN | PCTS | Base classes | Incremental classes | All classes |
|---|---|---|---|---|
| × | × | 91.65 | 88.85 | 91.43 |
| √ | × | 91.96 | 91.28 | 91.91 |
| × | √ | **92.19** | 93.42 | 92.29 |
| √ | √ | 92.16 | **97.91** | **92.62** |

## 4. Conclusions

In this work, we tried to solved a new problem of FCIAC using prototype adaptation and pseudo class-adjustable training. Experimental results on three public datasets show that our method outperformed state-of-the-art methods in average accuracy. In addition, both the proposed class-adjustable prototype adaptation network and pseudo class-adjustable training strategy have contributions to the performance improvement of our method.

The work in this paper is a preliminary work for addressing the FCIAC problem. Its performance still needs to be improved. In next work, we will consider optimizing the structure of the model and loss for further enhancing the performance of our method.

## 5. Acknowledgements

This work was supported by the national natural science foundation of China (62371195, 62111530145, 61771200), and the project of the10th Meeting of the China-Croatia Science and Technology Cooperation Committee (No. 10-34).

## 6. Generative AI Disclosure

We used GPT-5.2 to assist in polishing the manuscript.